\documentclass[pra,onecolumn,preprint]{revtex4-1}
\usepackage{amsmath,amssymb}
\usepackage{graphicx}

\def\p{{\bm{p}}}
\def\q{{\bm{q}}}
\newcommand{\be}{\begin{equation}}
\newcommand{\ee}{\end{equation}}

\newcommand{\vecp}{{\mathbf p}}
\newcommand{\vecq}{{\mathbf q}}

\newcommand{\x}{{\mathbf x}}

\newcommand{\K}{{\mathbf K}}

\newcommand{\opA}{{\hat{A}}}

\newcommand{\tr}{\textrm{Tr\,}}

\newcommand{\e}{\textrm{e}}

\newcommand{\vct}[1]{\ensuremath\mbox{\boldmath$ #1 $}}
\newcommand{\Vxi}{\vct \xi}
\newcommand{\Veta}{\vct \eta}

\def\ro{\hat{\rho}}
\begin{document}
\title{Local quantum ergodic conjecture}
\author {Eduardo Zambrano}
\email{zambrano@pks.mpg.de}
\affiliation{Max-Planck-Institut f\"ur Physik komplexer Systeme, 
N\"othnitzer Str. 38, 01187, Dresden, Germany  
}
\author{W.P. Karel Zapfe}
\affiliation{Centro Brasileiro de Pesquisas Fisicas,
Rua Xavier Sigaud 150, 22290-180, Rio de Janeiro, R.J., Brazil}
\author{Alfredo M. Ozorio de Almeida}
\email{ozorio@cbpf.br}
\affiliation{Centro Brasileiro de Pesquisas Fisicas,
Rua Xavier Sigaud 150, 22290-180, Rio de Janeiro, R.J., Brazil}

\begin{abstract}

The Quantum Ergodic Conjecture equates the Wigner function for a typical eigenstate of a classically
chaotic Hamiltonian with a delta-function on the energy shell. This ensures the evaluation
of classical ergodic expectations of simple observables, in agreement with Shnirelman's theorem,
but this putative Wigner function violates several important requirements. Consequently, 
we transfer the conjecture to the Fourier transform of the Wigner function, that is, the chord function. 
We show that all the relevant consequences of the usual conjecture 
require only information contained within a small (Planck) volume 
around the origin of the phase space of chords: translations in ordinary phase space.
Loci of complete orthogonality between a given eigenstate and its nearby translation
are quite elusive for the Wigner function, but our local conjecture
stipulates that their pattern should be universal for ergodic eigenstates 
of the same Hamiltonian lying within a classically narrow energy range. 
Our findings are supported by numerical evidence in a Hamiltonian exhibiting soft chaos.
Heavily scarred eigenstates are remarkable counter-examples of the ergodic universal pattern.
\end{abstract}

\maketitle
\section{Introduction}
The time average of any observable evolved by an ergodic classical Hamiltonian
is determined by the delta distribution over the energy shell. 
The Quantum Ergodic Conjecture (QEC) of Voros \cite{Voros:1976} and Berry \cite{Berry:1977b}
states that the same distribution can be reinterpreted as the semiclassical limit 
of the Wigner function \cite{Wigner:1932}, for an eigenstate of the corresponding quantum 
Hamiltonian that has the same eigenenergy. 

As a consequence of this conjecture, the expectation value of a 
quantum observable is close to a classical average in phase space. 
Indeed, Shnirelman's theorem \cite{Shnirelman:1974,Colin:1985,Zelditch:2006} 
confirms asymptotically the conjecture for almost all eigenstates, 
albeit with severe restrictions for the allowed classes both of the observables and of the ergodic systems.
(see e.g. Refs. \cite{Baecker_Schubert:1998,Anantharaman:2013} and references therein).
It should be noted that the QEC is stronger than Shnirelman's theorem, with consequences 
that are much more subtle than the gross
correspondence of expectations to their classical values. 
Most notable, so far, is the deduction by Berry \cite{Berry:1977b} 
of local statistical correlations of wave functions, that are directly applied 
to the interpretation of experiments in quantum dots (see Ref. \cite{Molina:2012} and references therein).
The ever increasing experimental refinement in manipulating quantum states
may allow one to access other delicate properties, such as interference phenomena,
for eigenstates of classically ergodic Hamiltonians. Thus, a more refined quantum
ergodic conjecture with further predictions within a more precise framework may
become a practical necessity.  

There are several features of a quantum state, specially a pure one,
that are not correctly portrayed by the Wigner function proposed by the QEC. 
They were recently reviewed in \cite{Ozorio:2014},
where it was shown that the operator that it represents is not even positive.
Here we focus on the behavior of quantum states subject to a phase space translation.
Its effect on the QEC Wigner function varies radically, depending on the choice
of equivalent formulas that can be employed to describe it, pointing to yet 
another inadequacy of the QEC.

Thus, we employ the QEC Wigner function as a stepping stone to formulate 
an alternative Local Quantum Ergodic Conjecture (LQEC) for the 
Fourier transform of the Wigner function,
in a neighborhood of the origin. 
LQEC encapsulates all desirable classical properties
of the standard QEC for the Wigner function.
No assumption needs to be made outside this region, so there are no pitfalls. 
Furthermore, it predicts an universal pattern for the overlap of eigenstates,
within a classically small energy window of a given ergodic Hamiltonian, 
with their translation by a small displacement. 

This paper is organized as follows: In section \ref{sec:LQEC},
after some preliminary definitions, the LQEC is presented.
In section \ref{sec:wfcorr}, under a restriction on the Hamiltonian,
we show that the LQEC implies the same results as the QEC for the wave function correlations 
for ergodic states. In section \ref{sec:Universal}, the universal pattern of
blind spots for ergodic states is presented, and tested in the Nelson
Hamiltonian.
Finally, we discuss our results in section \ref{sec:Discussion}.
The semiclassical theory near to the origin is summarized in the Appendix \ref{ap:short}.
Intermediary steps for the wave function correlations are contained in the Appendix \ref{ap:wfcorr}.

\section{LQEC and small chords\label{sec:LQEC}} 
Denoting by $\x=(\vecp,\vecq)\in\mathbb{R}^{2D}$ a phase-space point,
the Wigner function \cite{Wigner:1932}, $W(\x)$, and its Fourier transform 
(known as the \emph{chord function} \cite{Ozorio:1998}), 
$\chi(\Vxi)$,
are both complete representations of the density operator, $\hat{\rho}$ \cite{Ozorio:1998}. 
They are conveniently defined
in terms of the unitary {\it translation operator}
\begin{equation}
\hat{T}_{\scalebox{.75}{$\Vxi$}} = 
e^{i\scalebox{.75}{$\Vxi$}\wedge \hat{\x}/\hbar}
=
\exp{\left[\frac{i}{\hbar}(\Vxi_{\vecp}\cdot \hat{\vecq}-\Vxi_{\vecq}\cdot \hat{\vecp})\right]} \ ,
\end{equation}
and in terms of the Fourier transform of 
$\hat{T}_{\scalebox{.75}{$\Vxi$}}$, the {\it reflection operator} $\hat{R}_{\x}$, 
such that
\begin{equation}
\label{eq:chordfunction}
\chi(\Vxi) = \frac{1}{(2\pi\hbar)^D}{\rm tr} \;(
\hat{T}_{-\scalebox{.75}{$\Vxi$}}
\;\hat{\rho})
,
\end{equation}
and
\begin{equation}
W(\x) = \frac{1}{(\pi\hbar)^D}{\rm
tr}\;(\hat{R}_{\x}\;\hat{\rho})
.
\label{eq:Wtr}
\end{equation}

According to the QEC,
the Wigner function that represents a typical $E$-energy eigenstate of a classically chaotic Hamiltonian
can be approximated by the classical {\it Liouville probability density}
for the $E$-energy shell of the Hamiltonian with Weyl symbol $H(\x)$
\footnote {$H(\x)$ coincides with the classical Hamiltonian, $H_C(\x)$, in the important case where
$H(\x)={\vecp}^2/2m + V(\vecq)$. For general mechanical observables, $\opA$, the approximation for the Weyl symbol, $A_C(\x)\approx A(\x)$, only holds asymptotically in the semiclassical limit.},
\begin{equation}
W_{\pi}(\x) = \frac{\delta(E-H(\x))}{\int d\x ~\delta(E - H(\x))},
\label{eq:ergodic}
\end{equation}
where $\delta$ denotes the Dirac delta-function.
The assumption is that the classical Hamiltonian is {\it ergodic}, so that the time average of any
classical observable, $A_C(\x)$, evolved by $H(\x)= H(\vecp,\vecq)$ according to Hamilton's equations,
coincides with the classical phase space average,
\begin{equation}
\langle \opA \rangle = \int d\x \;W(\x)\;A(\x) \ ,
\label{eq:expectation}
\end{equation}
for the classical probability density \eqref{eq:ergodic}. The conjecture is then that 
one can reinterpret this same ergodic distribution, $W_{\pi}(\x)$, 
as the approximate quantum Wigner function that represents the $E$-eigenstate of the corresponding quantum Hamiltonian.

Let us focus on the behavior of quantum states subject to a translation.
The overlap of a pure state with its translation is
\begin{equation}
|\langle \psi|\psi_{\scalebox{.75}{$\Vxi$}}\rangle|^2
=
|\langle \psi|\hat{T}_{\scalebox{.75}{$\Vxi$}}|\psi\rangle|^2
= (2\pi\hbar)^D \int d\x~ W(\x + \Vxi)~ W(\x),
\label{eq:Wigcor}
\end{equation}
so that in the case of \eqref{eq:ergodic} this would decrease monotonically to zero, until there is no more
overlap between the energy shell and its translation. However, this scenario is contradicted 
by an alternative expression for the expectation of the translation of a pure state, that is,
\begin{equation}
\chi(-\Vxi) \equiv \langle\hat{T}_{\scalebox{.75}{$\Vxi$}}\rangle = 
\int \frac{d\x}{(2\pi\hbar)^D} \;W(\x)\; 
e^{-i\scalebox{.75}{$\Vxi$}\wedge \x/\hbar}
= \langle \psi|\psi_{\scalebox{.75}{$\Vxi$}}\rangle.
\label{eq:chift}
\end{equation}
Indeed, insertion of \eqref{eq:ergodic} into this formula reveals a fine structure of oscillations
very close to the maximum at the origin of the phase space of chords, $\Vxi$, including
zeroes, i.e. zero overlap between the state and its translation.

This apparent paradox is resolved by shifting of the QEC conjecture for the Wigner function 
to the LQEC conjecture for the chord function:
\begin{equation}
\chi_E(\Vxi) \rightarrow 
\int \frac{d\x}{(2\pi\hbar)^D} \;W_\pi(\x)\;
e^{-i\scalebox{.75}{$\Vxi$}\wedge \x/\hbar}
.
\label{eq:LQEC}
\end{equation}
Here, both limits, $\Vxi\rightarrow 0$ and $\hbar\rightarrow 0$, should be interpreted 
in a heuristic sense. This is familiar in the use of the semiclassical limit, in which
good approximations may follow from the extrapolation to finite values of Planck's constant
of results derived in the strict asymptotic limit, $\hbar \rightarrow 0$. It will be shown
that such extrapolations also hold for the short chord limit: The region in which the LQEC
is presumed to hold is a Weyl volume of order $\hbar^D$ surrounding the origin $\Vxi=0$.

The expression of the chord function in terms of its moments 
that is presented in the Appendix A implies that all moments calculated
from the LQEC are classical, as are the expectation values of any observable
that is polynomial in the components of position and momentum.
Furthermore, the first few moments allow for a rough
estimate of the location of the zeroes of $\chi(\Vxi)$ close to the origin, 
generalizing the discussion in Ref. \cite{Zambrano_Ozorio:2010}.
It is shown that there is a nodal surface for the imaginary part of $\chi(\Vxi)$
that contains the origin and is locally parallel to $\langle\hat\x\rangle$.
On the other hand, the lowest quadratic moments, forming the covariance matrix \cite{Schrodinger:1930}, $\K$,
lead to an approximate nodal surface for the real part of the chord function
in the form of the little ellipsoid, namely 
\begin{equation}
\langle(\Vxi\wedge\hat \x)^2\rangle = \Vxi~J^{\mathsf{T}}\mathbf K J~ \Vxi = 2\hbar^2,
\label{eq:ellipse}
\end{equation} 
where $J$ is the standard symplectic matrix and $\mathsf{T}$ stands for the transpose.
These codimension-1 nodal surfaces intersect in codimension-2 surfaces,
which have zero dimension in any two dimensional section of the chord function,
so they generalize the {\it blind spots}, studied in Ref. \cite{Zambrano_Ozorio:2010} 
for the special case where $D=1$.

\section{Wave function correlations\label{sec:wfcorr}}
So far, no restriction has been made on the momentum dependence of the Hamiltonian, but the second order 
Schr\"odinger equation follows from quadratic momenta, i.e. the Hamiltonian separates 
into a quadratic momentum term and a potential energy,
\footnote{Note that masses, $m_j$, can be absorbed by the symplectic transformation,
$p'_j \mapsto \sqrt{2 m_j}~p_j$ and $\sqrt{2 m_j}~q'_j \mapsto q_j$, for each pair of conjugate
phase space coordinates.}
\begin{equation}
H(\x) = \vecp^2 + V(\vecq).
\label{quadp}
\end{equation}
Then, the approximate chord function near the origin is
\begin{equation}
\chi_E(\Vxi) 
\approx 
\frac{1}{(2\pi\hbar)^D}
\int d\vecq~ \e^{-\frac{i}{\hbar}\scalebox{.75}{$\Vxi$}_{\vecp}\cdot~{\vecq}} ~
\int d\vecp~ \e^{\frac{i}{\hbar}\scalebox{.75}{$\Vxi$}_{\vecq}\cdot~{\vecp}}~ 
\frac{\delta(E-\vecp^2 - V(\vecq))}{\int d\vecp~ \delta(E-\vecp^2 - V(\vecq))}, 
\label{eq:LQEC2}
\end{equation}
so that, integrating over the surface of the $\vecp$-sphere with radius $r_E(\vecq)\equiv\sqrt{E-V(\vecq)}$, 
one obtains
\begin{equation}
\chi_E(\Vxi) 
\approx \frac{1}{(2\pi\hbar)^D
~\Omega_{\vecq}}
\int d\vecq~ \e^{-\frac{i}{\hbar}\scalebox{.75}{$\Vxi$}_{\vecp}\cdot~{\vecq}} 
~F_D (|\Vxi_{\vecq}|; \vecq, E),
\label{eq:LQEC3}
\end{equation}
where we define
\begin{equation}
F_D (|\Vxi_{\vecq}|; \vecq, E) \equiv \Gamma(D/2)~
\frac{{\rm J_{\frac{D}{2}-1}}\left[r_E(\vecq)|\Vxi_{\vecq}| /{\hbar}\right]}
{\left[r_E(\vecq)|\Vxi_{\vecq}| /{2\hbar}\right]^{\frac{D}{2}-1}}
,
\label{eq:F}
\end{equation}
and $\chi_E(\Vxi)=0$ otherwise.
Here, ${\rm J}_{\nu}(x)$ denotes the Bessel function of 
order $\nu$ \cite{Abramowitz:1964}. 
Then, the $D$-volume for the region $V(\vecq)< E$ is just
\begin{equation}
\Omega_{\vecq}= \int d\vecq ~F_D (0; \vecq, E),
\end{equation}
since $F_D (0;\vecq, E)=1$.
For $D=2$, one has simply $F_D (|\Vxi_{\vecq}|; \vecq, E)={\rm J}_0(r_E(\vecq)|\Vxi_{\vecq}|/{\hbar})$. 

The chord function, within the plane $\Vxi_{\vecq}=0$, 
is merely the Fourier transform of the characteristic function for the region $V(\vecq)< E$, 
divided by the $D$-volume of this region. In the special case of a billiard, 
for which $V(\vecq)=0$ within its boundaries, there is no $\vecq$-dependence in $F$, 
so that it factors out of \eqref{eq:LQEC3}. Then, the dependence of the chord function 
on $\Vxi_{\vecq}$ resides only in $F$, divided by the $D$-volume of the billiard, 
and this is multiplied by the $\Vxi_{\vecp}$-dependent Fourier transform 
of the characteristic function for the interior of the billiard. 

The local chord function \eqref{eq:LQEC3} is sufficient for the derivation of Berry's (local) 
wave function correlations \cite{Berry:1977b}, defined as
\begin{equation}
{C}_{\Delta}(\Vxi_{\vecq}, {\mathbf Q}) 
\equiv  
\frac{1}{\mathcal{N}_\Delta(\mathbf{Q})}
\int \frac{d\vecq}{(2\pi\Delta)^{D/2}} 
\e^{-\frac{(\vecq-{\mathbf Q})^2}{2{\Delta}^2}} ~
\langle \vecq + \Vxi_{\vecq}/{2}|\psi\rangle
\langle \psi|\vecq - \Vxi_{\vecq}/{2}\rangle
,\label{eq:wavecor1}
\end{equation}
where $\mathcal{N}_\Delta(\mathbf{Q})$ is a normalization factor such that $C_\Delta(0,\mathbf{Q}) = 1$.
This correlation function can be obtained as a Fourier transform of the chord
function in a small neighborhood of the origin, as shown in the Appendix \ref{ap:wfcorr}.

For the case of ergodic eigenstates, one inserts 
the approximate chord function of the LQEC \eqref{eq:LQEC} into \eqref{eq:wavecor2},
leading to
\begin{equation}
C_{\Delta}(\Vxi_{\vecq}, {\mathbf Q}) 
\approx
\frac{1}{{\mathcal N}_{\Delta}(\mathbf Q)} 
\int\hspace{-.1cm} \frac{d\x}{(2\pi\hbar)^D}~ 
\e^{i\scalebox{.75}{$\Vxi$}_{\vecq}\cdot \vecp/\hbar} ~
\frac{\delta(E-H(\x))}{\int d\x ~\delta(E - H(\x))}
\int\hspace{-.1cm} d\Vxi_{\vecp}~ 
\e^{i\scalebox{.75}{$\Vxi$}_{\vecp}\cdot ({\mathbf Q}- \vecq)/\hbar
-\Delta^2\scalebox{.75}{$\Vxi$}_{\vecp}^2/2{\hbar}^2},
.
\label{eq:wavecor3}
\end{equation}

Then, assuming the Hamiltonian in the form \eqref{quadp}, 
so that the chord function is given by \eqref{eq:LQEC3}, 
the correlation is given directly by \eqref{eq:wavecor2} as
\begin{equation}
{C}_{\Delta}(\Vxi_{\vecq}, {\mathbf Q}) \approx
\frac{\int d\vecq~ \e^{-\frac{({\mathbf Q}- \vecq)^2}{2{\Delta}^2}}~F_D (|\Vxi_{\vecq}|; \vecq, E)}
{\int d\vecq~ \e^{-\frac{({\mathbf Q}- \vecq)^2}{2{\Delta}^2}}~F_D (0; \vecq, E)}
\approx \frac{F_D (|\Vxi_{\vecq}|; {\mathbf Q}, E)}{F_D (0;{\mathbf Q}, E)}
\label{wavecor4}
\end{equation}
where the last approximation holds under the assumption that $\Delta$ is classically small, 
so that one can
neglect the variation in $V(\vecq \approx {\mathbf Q})$. This is certainly the case for the local classical approximation of the chord function within a Planck volume. No assumption about the outlying regions of the chord function is required, so as to re-derive the isotropic wave function correlations
of \cite{Berry:1977b}. In the limit of a billiard, i.e $V(\vecq)=0$, 
the averaging region is not constrained, so that local correlations 
can be identified with the global correlation for the wave function. 

\section{The universal local chord function\label{sec:Universal}}
Our general assumption is that the chord function for an energy eigenstate 
within a Planck volume of the origin is determined by corresponding classical structures 
at its eigenenergy. In the case of an integrable Hamiltonian of a system with two
or more degrees of freedom, neighboring states correspond to quantized tori that
need not lie close to each other within their neighboring energy shells.
Hence, the local chord functions for neighboring states may be quite diverse.
This will also hold for the pattern of their zeroes, that is, loci of possible translations
of each eigenstate, where it becomes orthogonal to itself.
The LQEC predicts an alternative scenario for the displacements of ergodic eigenstates:
They all share a common classical structure, the full energy shell itself, and hence,
for each Hamiltonian, the loci of zeroes of their local chord functions is universal, 
within a classically narrow energy range. 
Sweeping through two dimensional sections of the codimension-2 manifolds of zeroes,
these zeroes arise generically as blind spots, just as for a single degree of freedom;
the pattern of these blind spots is then predicted to be universal for neighboring
ergodic states. Furthermore, the pattern of nodal lines for the real and the imaginary part
of the local chord function should be reasonably approximated by Eq. \eqref{eq:LQEC}.

Any averaging over the chord function, or the Fourier transform of the Husimi function,
masks the pattern of its zeroes. Therefore, the best way to visualize them is by
examining two dimensional sections of the phase space of chords, that is, the locus of
all possible translations. Then, the zeroes are typically isolated points of the section,
{\it blind spots} just as in the case $D=1$, lying at the intersection of the nodal lines
of the real and the imaginary parts of the chord function \cite{Zambrano_Ozorio:2010}. 

In the case that the potential energy, and hence the Hamiltonian, has a center of (reflection)
symmetry, the chord function is real, and hence the zeroes in any section are identified
with the nodal lines of the real chord function. Moreover, the chord function is then
related by a trivial re-scaling to the Wigner function itself \cite{Ozorio_etal:2005}. Considering that a reflection
symmetry is in no way incompatible with classical chaos, this is a dramatic falsification
of a fully detailed account of the QEC for the Wigner function, which lacks the highest 
peak of any symmetric Wigner function at the reflection center (the origin), 
together with the surrounding oscillations. In contrast, this is precisely the region to which the LQEC applies
and is here studied numerically.

Note that Eq. \eqref{eq:LQEC} can be evaluated numerically by Monte Carlo integration as
\begin{equation}
\chi_E(\Vxi)\approx \frac{1}{N}\sum_{j=1}^N \e^{-i\x_j\wedge\scalebox{.75}{$\Vxi$}/\hbar}
\quad\quad (N\to\infty)
\end{equation}
where $\x_j$ are random phase-space points sampled according to $W_\pi(\x)$.
In practice, the sampling is performed using a sufficiently narrow Gaussian about the energy-shell, 
namely
\begin{equation}
W_\pi(\x) \to\frac{1}{\sigma\sqrt{2\pi}}\exp\left[-\frac{(H(\x)-E)^2}{2\sigma^2}\right]
\quad\quad
\,\, (\sigma\to0^+).
\end{equation}

We will consider the classical Nelson Hamiltonian is:
\begin{equation}
\label{eq:Nelson}
H({\bf x}_1,{\bf x}_2)\equiv
\frac{(p_1^2+p_2^2)}2 + \frac{q_1^2}2+
\left(q_2-\frac{q_1^2}2\right)^2
\;,
\end{equation} 
and its quantum counterpart results from the replacement: 
${\bf x}_1\equiv(q_1,p_1)\rightarrow (\hat{q_1},\hat{p_1})$ and
${\bf x}_2\equiv(q_2,p_2)\rightarrow (\hat{q_2},\hat{p_2})$ in (\ref{eq:Nelson}).
The classical dynamics of this system is mixed for all energies, but we consider 
eigenstates $|\psi_n\rangle$ whose eigenenergies correspond to an essentially 
chaotic classical behavior (see Ref. \cite{Prado_Aguiar:1994} and references therein). 

The exact quantum calculations are performed using the 
imaginary-time nonuniform mesh method (ITNUMM) proposed 
by Hernando et al. \cite{Hernando:2013}. This method is 
specially designed for the calculation high-excited states 
and it has a very easy implementation.

\begin{figure}
\centering
\includegraphics[width=0.75\textwidth]{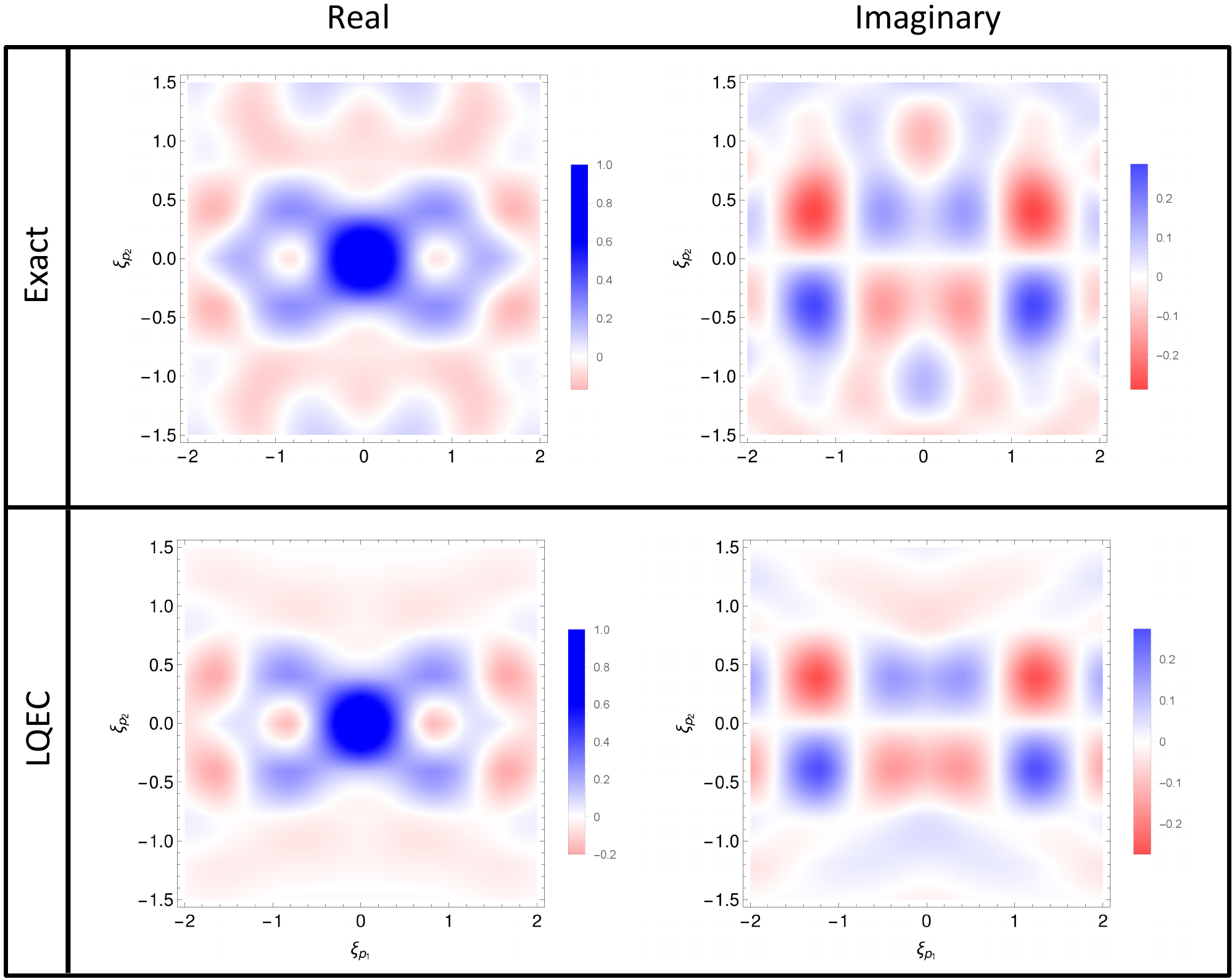}
\caption{\label{fig:n109}
Exact quantum (upper) and semiclassical LQEC (lower) chord functions 
for the $109^{\rm th}$ state of the Nelson Hamiltonian \eqref{eq:Nelson} in the 
$\Vxi_{\vecp}$-plane.
Here $\hbar=1$.
}
\end{figure}

\begin{figure}
\centering
\includegraphics[width=0.75\textwidth]{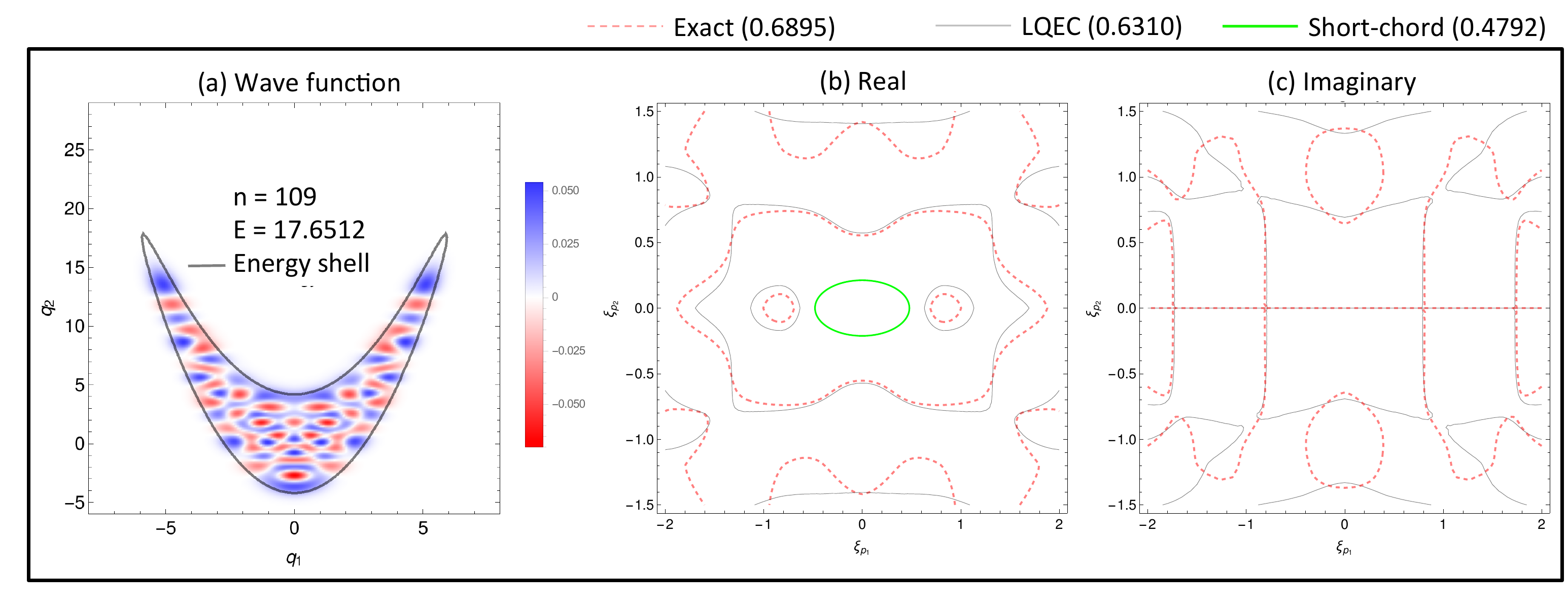}
\caption
{
\label{fig:109nl}
109$^{\rm th}$ excited eigenstate of the Nelson Hamiltonian \eqref{eq:Nelson}.
(a) Exact wave function. (b) Real and (c) imaginary nodal lines of the chord function 
in the $\Vxi_{\vecp}$-plane, evaluated by several methods: 
exact quantum (red dashed), semiclassical LQEC (gray solid) and short-chord ellipse (green solid).
In parenthesis is the estimation of the first blind spot along the horizontal axis for each method.
Here $\hbar=1$.
}
\end{figure}

\begin{figure}[htb!]
\centering
\includegraphics[width=0.75\textwidth]{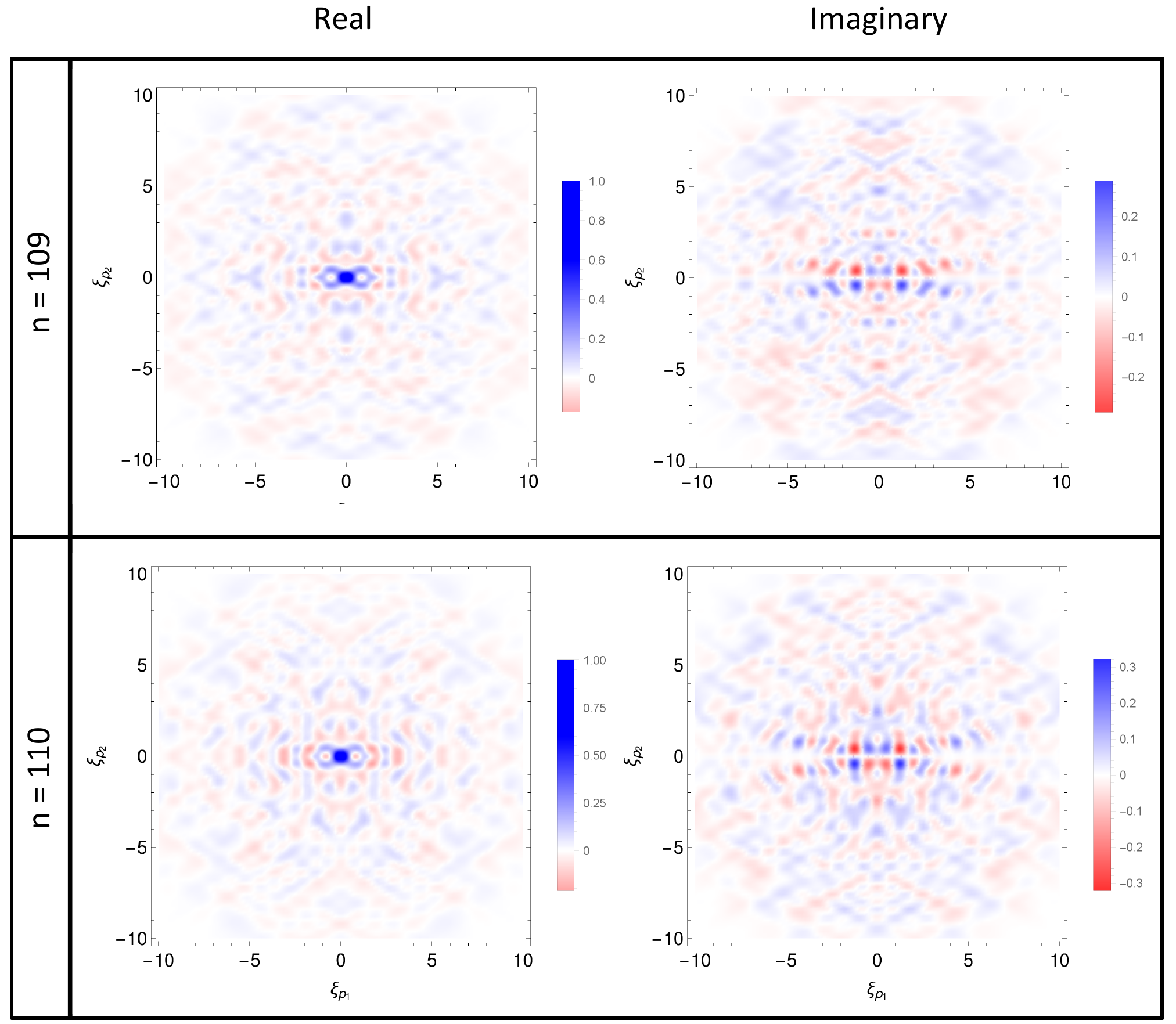}
\caption{\label{fig:qm}
Exact chord functions for the $109^{\rm th}$ and $110^{\rm th}$ excited state of the 
Nelson potential in the $\Vxi_{\vecp}$-plane. Here $\hbar=1$. 
}
\end{figure}

For simplicity, we restrict to the plane of the momenta $\Vxi_{\vecp}$, defined by $\Vxi_\vecq=0$.
In a small (Planck) region about the origin, Fig. \ref{fig:n109} shows 
the LQEC and the exact chord function for the $109^{\rm th}$ excited state of the 
Nelson Hamiltonian in the plane of momenta.
We observe that the LQEC furnishes an adequate description of the chord 
function in this small region about the origin. The blind spots arise as intersections of 
the nodal surfaces of the real and imaginary parts of $\chi(\Vxi)$ in the 
appropriate section. 
A comparison between of the exact and the LQEC nodal lines of the 
chord function for $109^{\rm th}$ state
are shown in Figs. \ref{fig:109nl}(b) and (c).  
The closest blind spots to the origin can be roughly 
estimated by the intersection of the 
the short-chord ellipsoid \eqref{eq:ellipse}
and the plane $ \Vxi \wedge\langle\hat\x\rangle = 0$. 
In the $\Vxi_{\vecp}$-plane, the latter is the horizontal axis, whereas as the former is 
depicted by the green line in Fig. \ref{fig:109nl}(b).
Figures \ref{fig:qm} show the exact quantum chord functions 
for two consecutive states, the $109^{\rm th}$ and the $110^{\rm th}$, respectively. 
Here a universal structure in a neighborhood of the origin is observed;
however, it is not present in regions far from the origin.
The chord function in \eqref{eq:LQEC} decay exponentially in the outer region.
This pattern depends only on the shape of the energy-shell.

\begin{figure}[hb!]
\centering
\includegraphics[width=0.75\textwidth]{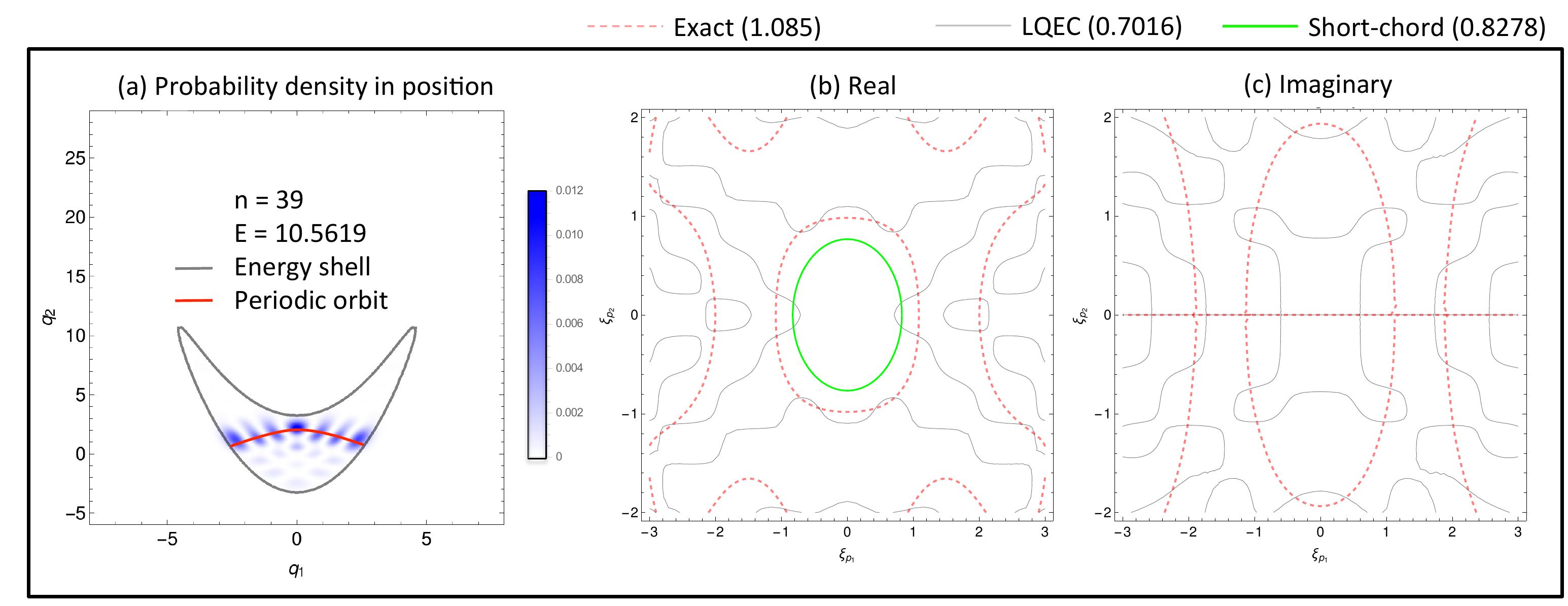}
\caption{\label{fig:scar}
Scarred state of the Nelson potential for the $39^{\rm th}$ state: 
(a) The density probability in position space $(q_1,q_2)$. Red line shows the classical periodic orbit. 
(b) Real and (c) imaginary nodal lines of the chord function: exact quantum (red dashed), semiclassical LQEC
(gray solid) and short-chord ellipse (green solid)
}
\end{figure}

For scarred states, the LQEC and the exact quantum differs substantially. 
Fig. \ref{fig:scar}(a) shows the $39^{\rm th}$ excited state of the Nelson 
potential and the respective scarring classical orbit. 
We observe that the nodal pattern of this scarred state differs noticeably from that of
typical ergodic states at nearly the same energy, so that the nearest blind spots
are displaced, as show in Figs. \ref{fig:scar}(b) and (c). 

\section{Discussion\label{sec:Discussion}}
The subtle nature of the transition from the QEC to the LQEC may benefit from an alternative view point:
The hard core of quantum ergodicity is Shnirelman's theorem and so concerns expectations of
mechanical observables, whose Weyl symbols are smooth functions. For this end the QEC is perfectly
adequate, even though one cannot consider Eq. \eqref{eq:ergodic} to be a proper Wigner function.
By including the real and the imaginary part of the chord function for small chords as the
expectations of the Hermitian operators (see Eq. \eqref{eq:cs} in the 
Appendix \ref{ap:short}) as legitimate observables for the QEC,
we substantially increase the predictive value of our ergodic conjecture, the LQEC, by encompassing
quantum features that are more refined than simple expectation values. This small region
of the chord function allows for the re-derivation of wave function correlations that were previously known
and predicts blind spots for small chords that are quite problematic for the QEC.

Perhaps, the greatest advantage is that the LQEC is not burdened by a full prediction that cannot be
dismembered within the Wigner picture. On the contrary, it is just within a tiny region of the
chord phase space that the ergodic chord function is determined, so that one avoids the violation
of general rules. By the identification of the LQEC with the small chord approximation \cite{Zambrano_Ozorio:2010}, 
in the case of systems with a single degree of freedom, one obtains a clear notion of the complete
irrelevance of the extrapolation of the LQEC beyond the Planck volume that it legitimately describes. 

The value of the outer region of the chord function is to reveal the purely quantum features
that distinguish quantum states. In the case of neighboring ergodic eigenstates,
the expression
\begin{equation}
|\langle \psi_n|\psi_{n'}\rangle|^2= (2\pi\hbar)^D \int d\Vxi~ \chi_n(\Vxi)~ \chi_{n'}(\Vxi)^*=0,
\label{chordoverlap}
\end{equation}
reveals that the similarity of both chord functions close to the origin must be compensated
by their differences further off. These differences, 
which can be discerned in Fig. \ref{fig:qm}, then highlight the gap in our knowledge of ergodic eigenstates.

\acknowledgements
It is a pleasure to thank Fabricio Toscano and Alberto Hernando de Castro for advice with the 
numerical calculations. 
Partial financial support from the National Institute for Science and Technology: Quantum Information, FAPERJ and CNPq is gratefully acknowledged.

\appendix

\section{Moments of position and momentum\label{ap:short}}

The definition of the chord function (\ref{eq:chift}) allows us to calculate the distribution moments, or statistical moments, of $\hat\p$ and $\hat\q$ in the form of the derivatives of the chord function, i.e. explicitly
\begin{equation}
\langle\hat{\vecp}^n\rangle=\tr\,\hat{\vecp}^n\ro
=(-i\hbar)^{n}\left.\frac{\partial^n\chi}{\partial\xi_{\vecq}^n}\right|_{\scalebox{.75}{$\Vxi$}=0}
,
\label{eq:deri-moment_p}
\end{equation}
and
\begin{equation}
\langle\hat{\vecq}^n\rangle=\tr\,\hat{\vecq}^n\ro
=(i\hbar)^{n}\left.\frac{\partial^n\chi}{\partial\xi_{\vecp}^n}\right|_{\scalebox{.75}{$\Vxi$}=0}.
\label{eq:deri-moment_q}
\end{equation}
Conversely, if we know all the moments, then we know the chord function, 
because the expansion in a Taylor series of the chord function is
\begin{equation}
\chi(\Vxi)=
\sum_{n=0}^{\infty}\frac1{n!}\sum_{k=0}^{n}\frac{(-1)^{k}}{(i\hbar)^n}\left(\hspace{-.2cm}
\begin{array}{c}n\\n-k\end{array}\hspace{-.2cm}\right)
\left\langle\mathcal{M}\left(\hat{\vecq}^{n-k}\hat{\vecp}^{k}\right)\right\rangle\hspace{.1cm}\xi_{\vecq}^k\xi_{\vecp}^{n-k},
\label{X-exp-momentdis}
\end{equation}
where, for conjugate components $(q_j, p_j)$
\begin{equation}
\mathcal{M}\left(\hat{\q_j}^{n}\hat{\p_j}^{k}\right)=
\frac{1}{n+k}\sum_{P_{nk}}\hat{\q_j}^{n}\hat{\p_j}^{k}
\label{permutations}
\end{equation}
and $P_{nk}$ stands for all possible permutations of products of $q^n$ and $p^k$. 
(No permutation is needed for commuting components.)
\par
Thus, knowledge of the chord function in an infinitesimal neighborhood of the origin
is equivalent to information on the expectation for all observables in the form of polynomials 
in position and momenta. 
In principle, the whole chord function  
would be determined by all statistical moments combined with analiticity,
this is no available in practice, due to the very oscillatory 
nature of highly excited eigenstates.

Recall that the chord function \eqref{eq:chift} is in general complex. 
Because it represents a Hermitian operator, then $\chi(-\Vxi) = \chi(\Vxi)^*$, 
where the asterisk denotes complex conjugation.
The real and imaginary parts of the chord function are related to the 
cosine and the sine transforms of the Wigner function, respectively.
By defining the Hermitian operators
$
{\hat c}_{\scalebox{.75}{$\Vxi$}}\equiv 
(
{\hat T}_{\scalebox{.75}{$\Vxi$}}+{\hat T}_{-\scalebox{.75}{$\Vxi$}}
)/
2,
$ 
and
$
{\hat s}_{\scalebox{.75}{$\Vxi$}}\equiv 
(
{\hat T}_{\scalebox{.75}{$\Vxi$}}-{\hat T}_{-\scalebox{.75}{$\Vxi$}}
)/2i
,
$ 
one can see that 
\begin{equation}
\mathfrak{Re}\chi(\Vxi)\propto
\langle\psi|{\hat c}_{\scalebox{.75}{$\Vxi$}}|\psi\rangle
\quad\text{ and }
\quad
\mathfrak{Im}\chi(\Vxi)\propto
\langle\psi|{\hat s}_{\scalebox{.75}{$\Vxi$}}|\psi\rangle
.
\label{eq:cs}
\end{equation}

\section{Small chords and wave function correlations\label{ap:wfcorr}}
Inserting the inverse chord transform \cite{Ozorio:1998}, 
\begin{equation}
\langle \vecq + \Vxi_{\vecq}/{2}|\psi\rangle\langle \psi|\vecq - \Vxi_{\vecq}/{2}\rangle 
= \frac{1}{(2\pi\hbar)^D} \int d\Vxi_{\vecp}~ \chi(\Vxi)
\e^{i\scalebox{.75}{$\Vxi$}_{\vecp}\cdot \vecq/\hbar},
\label{invcohrd}
\end{equation}
into the local wave function correlation \eqref{eq:wavecor1}, one obtains 
\begin{equation}
{C}_{\Delta}(\Vxi_{\vecq}, {\mathbf Q}) 
= \frac{1}{{\mathcal N}_{\Delta}(\mathbf Q)}
\int d\Vxi_{\vecp}~ 
\chi(\Vxi_{\vecq},\Vxi_{\vecp})~
\e^{i 
\scalebox{.75}{$\Vxi$}_{\vecp}\cdot {\mathbf Q}/\hbar} ~
\e^{-\Delta^2\scalebox{.75}{$\Vxi$}_{\vecp}^2/2\hbar^2}.
\label{eq:wavecor2}
\end{equation}
Note that the normalization constant can be expressed as 
\begin{equation}
{\mathcal N}_{\Delta}(\mathbf Q) = 
\int d\Vxi_{\vecp}~ \chi(0,\Vxi_{\vecp})~
\e^{
i\scalebox{.75}{$\Vxi$}_{\vecp}\cdot {\mathbf Q}/\hbar} ~
\e^{
-\Delta^2\scalebox{.75}{$\Vxi$}_{\vecp}^2/2{\hbar}^2}
.
\label{eq:NDelta}
\end{equation}

Let us then choose the smoothing parameter as $\Delta\approx {\hbar}^{1/2}$, i.e. the
width of a coherent state, so that the Gaussian integration window in Eq. \eqref{eq:wavecor1} 
has a volume $\hbar^{D/2}$. Within this classically small volume, the wave function 
for a highly excited eigenstate necessarily involves an average over many oscillations
of wave length $\hbar$. Then, the same Gaussian width is obtained for the smoothing 
of the chord function in \eqref{eq:wavecor2}, so that the local definition of the wave function correlation
only requires the information of the chord function within a Planck volume surrounding the origin.
It should be noted that the chord function of a state
that is translated in phase space by a chord $\Veta$,
\begin{equation}
\chi_{\scalebox{.75}{$\Veta$}}(\Vxi) = \chi(\Vxi)~ \e^{i\scalebox{.75}{$\Vxi$}\wedge \scalebox{.75}{$\Veta$}/\hbar},
\end{equation}
leads to the interpretation of \eqref{eq:wavecor2} as the Gaussian smoothed projection
of the chord function for the eigenstate translated by $\Veta = ({\mathbf Q}, 0)$. 

Consider the effect of a further smooth projection, but now onto the ${\mathbf P}$-axis
with the specific choice $\Delta = {\hbar}^{1/2}$:
\begin{equation}
\int \frac{d\Vxi_{\vecq}}{(\pi\hbar)^D}~ 
{\mathcal{N}}_{\Delta}(\mathbf Q)~{C}_{\Delta}(\Vxi_{\vecq},{\mathbf Q})~
\e^{-i\scalebox{.75}{$\Vxi$}_{\vecq}\cdot {\mathbf P}/\hbar
-{\Delta}^2\scalebox{.75}{$\Vxi$}^2_{\vecq}/2\hbar^2}
= \int \frac{d\Vxi}{(\pi\hbar)^D}
\e^{
-\scalebox{.75}{$\Vxi$}^2/2\hbar
-i\scalebox{.75}{$\Vxi$}\wedge {\mathbf X}/\hbar
}
    \chi(\Vxi)~ 
= {\mathcal H}(\mathbf{X}).
\label{Husimi}
\end{equation}
Thus, we identify the Husimi function \cite{Husimi:1940,Takashi:1986}, $\mathcal{H}$,
evaluated at the phase space point, ${\mathbf X}=({\mathbf Q,{\mathbf P}})$,
that may be defined as the Gaussian coarse-graining of the Wigner function.
Surprisingly, the product of a Gaussian amplitude on the chord function does not
diminish its information content, since the Husimi function is also a complete representation of a quantum state.
Here we find that exact knowledge of the local correlation for all positions, ${\mathbf Q}$ also contains, 
in principle, complete information about the state. In the case of the Husimi function, 
information is delicately encoded in its analytic properties and may be hard to extract.

\bibliography{biblio}
\end{document}